\documentclass[pra,twocolumn,preprintnumbers,amsmath,amssymb,superscriptaddress]{revtex4} 
\usepackage[dvipdfmx]{graphicx}
\usepackage{epsfig}
\usepackage{subfigure}
\usepackage{bm}
\usepackage{braket}
\usepackage[dvipsnames]{xcolor}
\usepackage{times}
\usepackage{amsmath,amsthm,amssymb}
\usepackage[colorlinks,bookmarks=false,citecolor=blue,linkcolor=red,urlcolor=blue]{hyperref}
\usepackage{physics}
\usepackage{graphicx}
\usepackage{ascmac}
\usepackage{mathrsfs}
\usepackage{color}
\usepackage{comment}
\usepackage[normalem]{ulem}

\makeatletter
\renewcommand\bibsection{%
  \let\addcontentsline\@gobblethree 
  \section*{\refname}%
}
\makeatother
\makeatletter
\renewcommand\bibsection{}
\makeatother

\newcommand{\im}{{\rm i}}

\usepackage{comment}
\usepackage{hyperref}
\usepackage{mathtools}
\usepackage[normalem]{ulem}
\hypersetup{colorlinks=true,citecolor=blue,linkcolor=blue,urlcolor=blue}
\usepackage{soul}

\newcommand{\be}{\begin{equation}}
\newcommand{\ee}{\end{equation}}
\newcommand{\bea}{\begin{eqnarray}}
\newcommand{\eea}{\end{eqnarray}}
\theoremstyle{plain}

\newtheorem*{thm*}{Theorem}

\theoremstyle{definition}

\newcommand\calC{{\cal C}}

\newcommand\calE{{\cal E}}
\newcommand\calF{{\cal F}}

\newcommand\calO{{\cal O}}

\newcommand\calS{{\cal S}}

\theoremstyle{definition}
\newtheorem{definition}{Definition}

\newcommand{\eq}[1]{\begin{align} #1 \end{align}}

\def\Tr{\mathop{\mbox{\normalfont Tr}}\nolimits}

\begin{document}

\setstcolor{red}
\title{Quantum Fisher Information as a Measure of \\Symmetry Breaking in Quantum Many-Body Systems
}
\author{Shion Yamashika}
\email{shion.yamashika@uec.ac.jp}
\affiliation{Department of Engineering Science, University of Electro-Communications, Chofu, Tokyo 182-8585, Japan}
 
\author{Shimpei Endo}
 \email{shimpei.endo@uec.ac.jp}
\affiliation{Department of Engineering Science, University of Electro-Communications, Chofu, Tokyo 182-8585, Japan}

 \author{Hiroyasu Tajima}
 \email{hiroyasu.tajima@inf.kyushu-u.ac.jp}
 \affiliation{Information Science and Electrical Engineering, Department of Informatics, Kyushu University, Nishi-ku, Fukuoka, 819-0395, Japan}
 \affiliation{JST FOREST, 4-1-8 Honcho, Kawaguchi, Saitama, 332-0012, Japan}

\maketitle
{\bf
Symmetry breaking underlies diverse phenomena from phase transitions in condensed matter to fundamental interactions in gauge theories. 
Despite many proposed indicators, a general quantification of symmetry breaking that is faithful, computable, and valid in the thermodynamic limit has remained elusive. Here, within quantum resource theory, we propose the quantum Fisher information (QFI) as such a measure. We demonstrate its utility by computing QFI for paradigmatic models: in the BCS superconductor, the QFI counts the number of Cooper pairs; in the transverse-field XY spin chains, it captures topological phase transition that has no local order parameter; and in quantum quench dynamics, it allows us to exactly derive the microscopic origin and conditions of the quantum Mpemba effect in terms of excitation propagation, including in the thermodynamic limit--beyond the reach of previous analyses.
Our results show that the QFI, which is a complete resource monotone in the resource theory of asymmetry that plays the role of entanglement entropy in entanglement theory, faithfully captures symmetry breaking in condensed-matter systems.
These results highlight the QFI as a universal and physically meaningful diagnostic of symmetry breaking in both equilibrium and non-equilibrium quantum many-body systems. 
} 
\\~\\ \noindent \textbf{\large Introduction}\\
Symmetry and its breaking underlie a broad spectrum of physical phenomena, from the unification of forces in the Standard Model to the emergence of superconductivity in condensed-matter systems. In many-body systems, symmetry governs phase transitions, determines the structure of low-energy excitations, and constrains the dynamics through selection rules and conservation laws. When investigating symmetry breaking, it is therefore essential to identify when it occurs and to what extent the symmetry is broken, in a manner comparable across models and platforms. 
\par 
Although numerous indicators of broken symmetry have been proposed to address this need, a definitive indicator that can serve as a common standard is still lacking. 
In the traditional Landau paradigm, local order parameters diagnose symmetry breaking~\cite{landau2013statistical}. While conceptually simple and physically intuitive, the Landau order parameters are not always sufficient: they do not provide a necessary and sufficient criterion for symmetry breaking (i.e., they are not faithful), and they may fail to capture non-local or geometric features of the system. More recently, entanglement asymmetry has been introduced as a measure of symmetry breaking rooted in quantum information theory~\cite{ares2023entanglement}. It is faithful and sensitive to system geometry, yielding significant insights in complex quantum many-body phenomena, most notably in studies of the quantum Mpemba effect~\cite{ares2025quantum,ares2023entanglement,rylands2024microscopic,murciano2024entanglement,turkeshi2025quantum,chalas2024multiple,yamashika2024entanglement,joshi2024observing}. However, entanglement asymmetry typically grows only logarithmically with system size~\cite{capizzi2024universal}, rendering it unsuitable for the thermodynamic limit, where symmetry-breaking phenomena—such as phase transitions—are rigorously defined. This gap is consequential: without a faithful and thermodynamically robust indicator, it is difficult to perform controlled finite-size scaling to the thermodynamic limit, to reliably diagnose symmetry breaking out of equilibrium, or to formulate operational statements—such as performance bounds for transport and response—that require a quantitative account of broken symmetry across models and platforms. We therefore require a measure that (1) is strictly positive if and only if the state breaks the symmetry (faithfulness), (2) is efficiently computable, and (3) remains nonzero and well-defined in the thermodynamic limit.
\par 
In this work, building on insights from resource theory, we propose that the quantum Fisher information (QFI) satisfies (1)-(3) and can be regarded as a standard indicator of symmetry breaking in quantum many-body systems. While the QFI is a core quantity in quantum metrology~\cite{helstrom1969quantum} and has recently been applied to probe quantum many-body phenomena such as dynamical phase transitions~\cite{guan2021identifying,munoz2023phase,zhou2023dynamical} and multipartite entanglement~\cite{pezze2009entanglement,hyllus2012fisher,frowis2012measures}, it has received comparatively little attention as an indicator of symmetry breaking. Nevertheless, it possesses all the desired properties of a symmetry-breaking measure: it is faithful, computable, physically meaningful, and well-defined in the thermodynamic limit. Furthermore, it serves as the ``complete'' resource measure in the resource theory of asymmetry (the resource theory treating symmetry and its breaking)~\cite{gour_resource_2008,Marvian_thesis,skew_resource,Takagi_skew,Marvian_distillation,marvian_operational_2022,YT,YT2,SMT,YMST}, just as entanglement entropy does in entanglement theory~\cite{RToE1}. From a resource-theoretic viewpoint, much as entanglement entropy standardised the quantification of entanglement across condensed-matter and high-energy physics, it is natural to anticipate that the QFI can serve as a standard diagnostic of symmetry breaking in these domains; the present work substantiates this expectation.
\par 
We substantiate this expectation by showing that the QFI is highly effective as a symmetry-breaking indicator in condensed-matter settings. Through analytical calculations in paradigmatic models, including Bardeen-Cooper-Schrieffer (BCS) superconductors, topological spin chains, and non-equilibrium quantum quenches, we demonstrate that the QFI faithfully captures symmetry breaking across these diverse physical models—even in the thermodynamic limit—and, as a result, reveals insights beyond conventional approaches. 
For instance, in the BCS superconductor the QFI incorporates the contribution of finite-temperature decoherence to symmetry breaking through the purity of the state, in the XY spin chain it exhibits non-analytic behaviour at the topological transition where no local order parameter exists, and in quantum quench dynamics of a spin chain it provides the approximation-free identification of the conditions and origins for the quantum Mpemba effect. 
In addition, these examples establish a consistent microscopic picture of symmetry breaking in which the QFI directly counts the number of Cooper pairs, highlighting its value as a complementary perspective on symmetry breaking in quantum many-body physics. 
\\~\\
\noindent {\bf \large Results}
\\
We begin by demonstrating that the QFI offers a theoretically well-founded and physically meaningful framework for characterising symmetry breaking. 
To this end, we first clarify its theoretical foundations within the resource theory of asymmetry, and then demonstrate its applicability through representative physical models in both equilibrium and non-equilibrium settings.
\\
{\bf Theoretical Foundation}
\\
Let $\rho$ be a quantum state and $X$ the generator of a U(1) symmetry transformation. 
The QFI associated with $\rho$ and $X$ is defined as 
\begin{align}
    F_\rho(X)=2 \sum_{i,j} \frac{(p_i-p_j)^2}{p_i+p_j} \abs{\bra{i}X\ket{j}}^2,
    \label{def:QFI}
\end{align}
where $\qty{p_i,\ket{i}}$ are the eigenvalues and eigenstates of $\rho$. 
The QFI quantifies the sensitivity of the state $\rho$ under infinitesimal unitary transformations generated by $X$, and vanishes when and only when $\rho$ is invariant under such transformations. 
The QFI therefore offers a natural means to quantify the degree to which a quantum state breaks the U(1) symmetry. 
\par 
The suitability of the QFI as a quantifier of symmetry breaking is justified by the resource theory of asymmetry, which treats symmetry breaking as a operational resource and characterises covariant operations that do not increase the resource (For details, see the supplementary Information). Within this framework, the QFI emerges as a resource monotone under U(1)-covariant operations, meaning that it cannot increase under any physical process respecting the symmetry described by the unitary representation generated by $X$. In addition to this monotonicity, the QFI is faithful, namely, it is strictly positive if and only if the state breaks the symmetry, and additive over tensor products of states, which implies extensivity with system size. This extensivity plays an important role in analysing many-body phenomena such as phase transitions and quantum thermalisation, where rigourous discussion often requires taking the thermodynamic limit. The QFI also encodes the non-local quantum correlations relevant to symmetry breaking, allowing it to capture phenomena that lack a conventional local order parameter~\cite{zhang2022multipartite,pezze2017multipartite,yin2019quantum}.
Beyond these properties, the QFI attains a distinguished status in the resource theory of asymmetry as it fully determines the optimal asymptotic conversion rate between states under U(1)-covariant operations. In this sense, it serves as a complete resource measure, analogous to the entanglement entropy in the entanglement theory or the nonequilibrium free energy in quantum thermodynamics. Such complete measures not only quantify how efficiently one resource state can be converted into another, but also provide the ordering structure of state transformations, thereby highlighting the QFI as a physically fundamental quantity.
\par 
For practical purposes, it is often convenient to express the QFI in alternative but equivalent forms. 
One such representation involves the Uhlmann fidelity between $\rho$ and its symmetry-transformed version: 
\begin{align}
    F_\rho (X) = -4 \partial_\theta^2 \mathcal{F}(\rho, e^{-\im \theta X}\rho e^{\im \theta X} )|_{\theta=0}, 
    \label{eq:QFI_Fidelity}
\end{align}
where $\mathcal{F}(\rho,\sigma)=\Tr_{} \sqrt{\sqrt{\rho}\sigma\sqrt{\rho}}$ denotes the Uhlmann fidelity. 
If $\rho$ is pure, $\rho=\ket{\psi}\bra{\psi}$, the QFI further simplifies  as
\begin{align}
    F_{\ket{\psi}} (X)=4V_{\ket{\psi}}(X), 
    \label{eq:QFI_pure}
\end{align}
with $V_{\ket{\psi}}(X)=\bra{\psi}X^2\ket{\psi}-\bra{\psi} X\ket{\psi}^2$ being the variance of $X$. 
From the expression~\eqref{eq:QFI_pure}, it is evident that the QFI reflects the fluctuations of the global operator $X$, thereby exhibiting extensivity and sensitivity to nonlocal correlations. 
Furthermore, for a general mixed state $\rho$, the QFI satisfies the convex-roof relation:
\eq{
F_\rho(X)=4\min_{\{q_l,\ket{\psi_l}\}:\rho=\sum_lq_l\ket{\psi_l}\bra{\psi_l}}\sum_{l}q_lV_{\ket{\psi_l}}(X)
}
where the minimization is taken over all pure state decompositions $\{q_l,\ket{\psi_l}\}$ of $\rho$, including non-orthogonal ones. This demonstrates that the QFI faithfully captures the quantum contribution to the fluctuations of $X$, namely, the part arising from quantum superposition.
\\
\noindent 
{\bf BCS Phase Transition}
\\
To demonstrate the utility of the QFI as a measure of symmetry breaking, we begin with the BCS phase transition~\cite{bardeen1957theory}. 
It describes a system of interacting fermions undergoing a transition from a normal phase with conserved particle number to a superconducting phase involving spontaneous U(1) particle-number symmetry breaking.
\par 
We consider the system of fermions described by the BCS Hamiltonian, 
\begin{align}
    H_\mathrm{BCS}
    = 
    \sum_{\mathbf{k},\sigma}(\xi_\mathbf{k}-\mu) c_{\mathbf{k}\sigma}^\dag c_{\mathbf{k}\sigma}
    -
    \sum_\mathbf{k}(c_{\mathbf{k}\uparrow}^\dag c_{-\mathbf{k}\downarrow}^\dag \Delta  +\mathrm{H.c.}),
    \label{eq:H_BCS}
\end{align}
where $c_{\mathbf{k}\sigma}~(c_{\mathbf{k}\sigma}^\dag)$ is the annihilation (creation) operator of a fermion with momentum $\mathbf{k}$ and spin $\sigma\in\qty{\uparrow,\downarrow}$, $\xi_\mathbf{k}$ denotes the single-particle dispersion, $\mu$ is the chemical potential, and $\Delta =U \sum_\mathbf{k}\langle c_{-\mathbf{k}\downarrow}c_{\mathbf{k}\uparrow}\rangle$ is the superconducting gap with an effective pairing interaction $U>0$. 
At zero-temperature, the wavefunction is given by the product form~\cite{schrieffer2018theory} 
\begin{align}
    \ket{\mathrm{BCS}}
    =
    \prod_\mathbf{k}
    \qty(u_\mathbf{k}+v_\mathbf{k}c_{\mathbf{k}\uparrow}^\dag c_{-\mathbf{k}\downarrow}^\dag)\ket{0}, 
    \label{eq:|Psi_BCS>}
\end{align}
with variational parameters $(u_\mathbf{k},v_\mathbf{k})$ determined so that the energy functional $\bra{\rm BCS}H_\mathrm{BCS}\ket{\rm BCS}$ takes minimum. If $v_\mathbf{k}\neq 0$, this state is not an eigenstate of the total-number operator $N=\sum_{\mathbf{k},\sigma}c_{\mathbf{k}\sigma}^\dag c_{\mathbf{k}\sigma}$, reflecting spontaneous breaking of the U(1) symmetry generated by $N$.  
\par 
The QFI with respect to $N$ for the pure BCS state~\eqref{eq:|Psi_BCS>} takes the form 
\begin{align}
    F_{\ket{\rm BCS}}(N)
    &=
    4V_{\ket{\rm BCS}}(N)
    =
    16 \sum_\mathbf{k} \abs{\langle  c_{-\mathbf{k}\downarrow}c_{\mathbf{k}\uparrow}\rangle }^2. 
    \label{eq:QFI_BCS_T=0}
\end{align}
Since each term $|\langle c_{-\mathbf{k}\downarrow}c_{\mathbf{k}\uparrow}\rangle|^2$ corresponds to the mode occupation number of Cooper pairs with momentum $\pm\mathbf{k}$, according to Eq.~\eqref{eq:QFI_BCS_T=0}, the QFI directly measures the total number of Cooper pairs. 
This result is completely in accordance with the standard Landau theory, where the superconducting gap $\Delta$ servers as the order parameter encoding the pair correlations. 
The QFI likewise captures these correlations, but, in addition, it provides direct interpretation of the symmetry breaking as  originating from the emergence of the Cooper pairs. 
\par 
The above analysis can be easily extended to finite temperature, where the system is described by a mixed thermal state $\rho_T\propto e^{- H_\mathrm{BCS}/T}$. 
In this case, the QFI can be calculated exactly using Eq.~\eqref{eq:QFI_Fidelity} together with Wick's theorem as (see Supplemental Information)
\begin{align}
    F_{\rho_T}(N)
    = 
    \sum_\mathbf{k} 
    \frac{16\abs{ \langle c_{-\mathbf{k}\downarrow}c_{\mathbf{k}\uparrow}\rangle}^2}{\sqrt{P_\mathbf{k}(T)}},
    \label{eq:QFI_BCS_finite_T}
\end{align}
where $P_{\mathbf{k}}(T)$ quantifies the purity of each mode, with $\prod_{\mathbf{k}} P_{\mathbf{k}}(T)=\Tr[\rho_T^2]$.
The expression \eqref{eq:QFI_BCS_finite_T} still represents the total number of Cooper pairs, but now includes the effect of thermal decoherence through the purity. 
This additional dependence on purity highlights a feature not captured by the conventional framework based on the order parameter, providing a more refined perspective on symmetry breaking and coherence at finite temperature. 
\par
\begin{figure}
    \raggedright
    \includegraphics[width=0.9\linewidth]{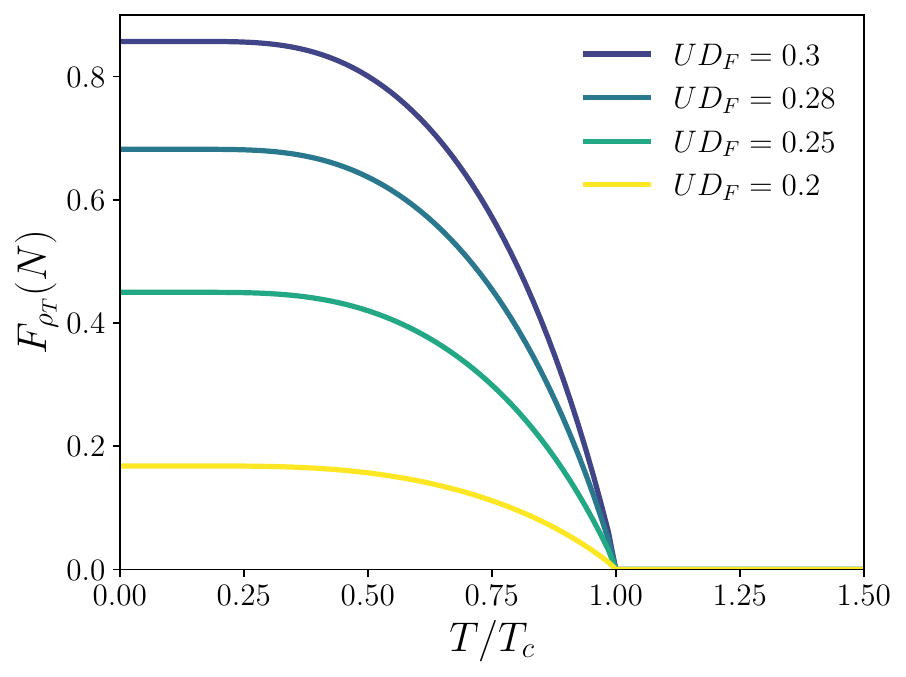}
    \caption{\textbf{Quantum Fisher information of the Gibbs ensemble of the BCS Hamiltonian~\eqref{eq:H_BCS}}. The curves are obtained by replacing the summation in Eq.~\eqref{eq:QFI_BCS_finite_T} with integral over energy introducing the density of states $D_F$ at Fermi level (see Supplemental Information).}
    \label{fig:QFI_BCS}
\end{figure}

 
Figure~\ref{fig:QFI_BCS} shows the temperature dependence of the QFI given in Eq.~\eqref{eq:QFI_BCS_finite_T}. The QFI remains finite in the superconducting phase and vanishes continuously at the critical temperature $T_c$, demonstrating the faithfulness of the QFI as an indicator of symmetry breaking. 
\\
\noindent \textbf{Topological Phase Transition}
\\
We have shown that the QFI serves as a faithful measure of spontaneous symmetry breaking in the BCS phase transition.  On the other hand, certain quantum phase transition, such as topological ones, occur without spontaneous symmetry breaking. 
In what follows, we show that the QFI remains a useful diagnostic tool even in such settings thanks to its sensitivity to long-range correlations that characterise non-local topological order. 
\par 
As a pedagogical example, we consider the transverse-field XY spin chain exhibiting a topological phase transition, described by a Hamiltonian 
\begin{multline}
    H_{\rm XY}
    = 
    -
    \sum_{i=1}^L
    \qty[
    \frac{1+\eta}{4} \sigma_i^x \sigma_{i+1}^x
    + 
    \frac{1-\eta}{4} \sigma_i^y \sigma_{i+1}^y
    +
    \frac{h}{2}\sigma_i^z
    ],
    \label{eq:H_XY}
\end{multline}
where $\sigma_i^{x,y,z}$ are the Pauli matrices acting on the $i$-th site, $h$ is the external transverse field, $\eta$ is the anisotropy parameter of the XY spin interaction, and $L$ is the system size.
This Hamiltonian is known to be equivalent to the Kitaev chain of a topological superconductor~\cite{kitaev2001unpaired}, and its ground state exhibits a quantum phase transition at $h=1$~\cite{kitaev2001unpaired,degottardi2011topological,lieb1961two}:
For $|h|<1$, the ground state belongs to a topologically non-trivial phase characterised by the presence of localized Majorana edge modes. In contrast, for $|h|>1$, the system is in a trivial phase without any topological feature. 
\par 
To probe the transition, we consider the QFI with respect to the global U(1) rotation around the $z$-axis, generated by the transverse magnetization $M = \sum_i \sigma^z_i / 2$. 
The QFI associated with $M$ is expected to serve as an indicator of the phase transition, by the fact that the ground state approximately respects this symmetry deep in the trivial phase ($|h|\gg1$), and increasingly breaks it as $|h|$ decreases toward the non-trivial phase.
\par 
The QFI can be analytically calculated by mapping the spins to fermions via the Jordan-Wigner transformation $c_i=(\prod_{j<i}\sigma_j^z)(\sigma_i^x+\im \sigma_i^y)/2$. With this mapping, the transverse magnetization becomes the total fermion number operator $M=\sum_{i}c_i^\dag c_i$ and the Hamiltonian~\eqref{eq:H_XY} transforms into a free fermion model. The corresponding ground state of this model reads~\cite{lieb1961two} 
\begin{align}
    \ket{g(h,\eta)} = \prod_{k>0}\qty(\cos\frac{\theta_k}{2} + \im \sin \frac{\theta_k}{2}c_k^\dag c_{-k}^\dag)\ket{0}, \label{eq:|g>}
\end{align}
where $\ket{0}$ is the fermionic vacuum state, $c_k=\sum_i e^{-\im ki}c_i/\sqrt{L}$, and the Bogoliubov angle $\theta_k$ is defined by $\tan\theta_k =\eta \sin k/(h-\cos k)$. 
Evaluating the variance of particle number in the state $\ket{g(h,\eta)}$, we obtain the QFI in the following form 
\begin{align}
    F_{\ket{g}}(M) = 8\sum_k \abs{\bra{g(h,\eta)}c_{-k}c_k\ket{g(h,\eta)}}^2. 
    \label{eq:QFI_XY_1}
\end{align}
Each term $\abs{\bra{g}c_{-k}c_k\ket{g}}^2$ in Eq.~\eqref{eq:QFI_XY_1} represents the number of Cooper pairs consisting of fermions with momenta $\pm k$, indicating that the QFI directly measures the total number of Cooper pairs, in formal analogy to the BCS case (see Eq.~\eqref{eq:QFI_BCS_T=0}). This result reflects the equivalence between the XY spin chain~\eqref{eq:H_XY} and the Kitaev chain of topological superconductors. 
\par 
Substituting Eq.~\eqref{eq:|g>} into Eq.\eqref{eq:QFI_XY_1} and taking the thermodynamic limit $L\to\infty$, the density of the QFI can be analytically calculated as 
\begin{align}
    f_{\ket{g}}(M)
    &=
    \lim_{L\to\infty}\frac{F_{\ket{g}}(M)}{L}
    \nonumber
    \\
    &=
    \begin{dcases}
        \frac{2\eta}{\eta+1} & \abs{h}\leq 1,\\
        \frac{2\eta^2}{1-\eta^2}\qty(\frac{\abs{h}}{h^2+\eta^2-1}-1)
        & \abs{h}>1. 
    \end{dcases}
    \label{eq:QFI_XY}
\end{align}
As expected, the QFI vanishes at $\eta=0$ or $|h|\to \infty$, at which the XY Hamiltonian~\eqref{eq:H_XY} preserves the U(1)-rotational symmetry around $z$ axis. 
\par
\begin{figure}
    \raggedright
    \includegraphics[width=0.9\linewidth]{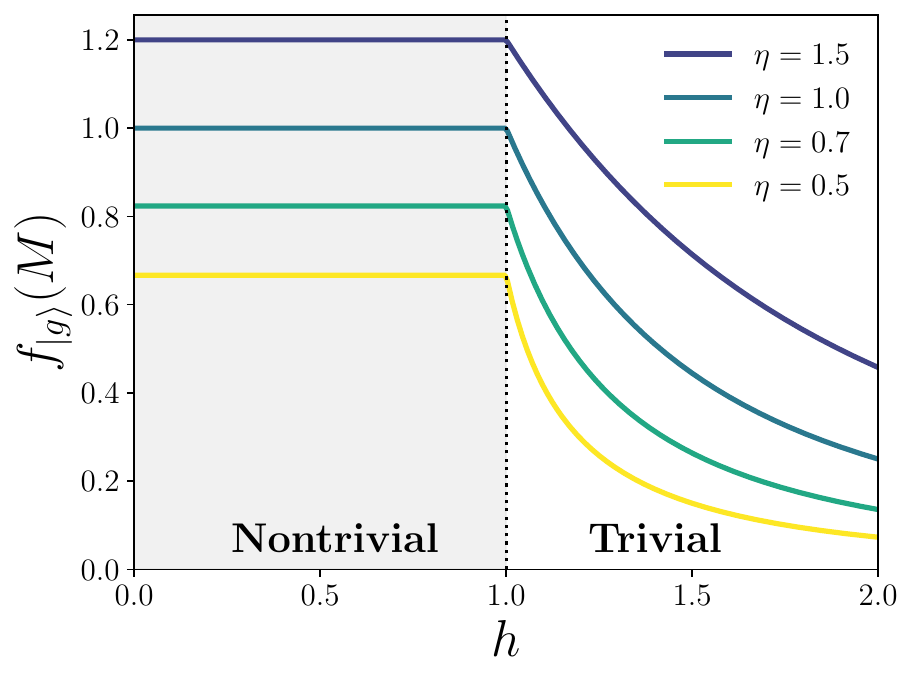}
    \caption{\textbf{Quantum Fisher information of the ground state of the XY Hamiltonian~\eqref{eq:H_XY}.} We plot the analytic expression~\eqref{eq:QFI_XY} as a function of $h$ for different values of $\eta$.}
    \label{fig:QFI_XY_GS}
\end{figure}

 
Figure~\ref{fig:QFI_XY_GS} shows the QFI as a function of $h$ with several fixed $\gamma$. 
It shows that the QFI remains constant in the topologically non-trivial phase ($h\leq 1$), whereas it monotonically decreases with increasing $h$ in the trivial phase ($h>1$). 
Notably, the QFI exhibits a non-analytic behaviour at the critical point $h=1$, clearly demonstrating that the QFI captures the topological transition. 
This ability of the QFI to precisely detect the topological transition stems from the fact that it incorporates information from all non-local pair correlations and remains finite in the thermodynamic limit.  
\\ 
\textbf{Quantum Mpemba effect in thermodynamic limit} 
\\
We now extend our analysis to a highly nonequilibrium phenomenon, the quantum Mpemba effect, in which a system prepared farther from equilibrium approaches its stationary state more rapidly than one initially closer to it. 
While various versions of this effect have been explored~\cite{nava2019lindblad,carollo2021exponentially,ares2023entanglement,chatterjee2023quantum,moroder2024thermodynamics,nava2024mpemba,strachan2025non,longhi2025mpemba,boubakour2025dynamical}, we focus here on isolated quantum systems, where the symmetry breaking plays a central role~\cite{ares2023entanglement}. 
\par 
To be more specific, we consider a quantum quench in the transverse-field XY chain~\eqref{eq:H_XY}. 
The system is initially prepared in the ground state $\ket{g(h,\eta)}$ of the Hamiltonian~\eqref{eq:H_XY}, which breaks the U(1) spin-rotational symmetry around the $z$-axis due to finite $h$ and $\eta$. 
At $t=0$, we abruptly set $h=\eta=0$, so that the system evolves unitarily as $\ket{\psi(t)}=e^{-\im t H_{\mathrm{XX}}}\ket{g(h,\eta)}$, where $H_\mathrm{XX}$ is the XX Hamiltonian obtained by inserting $\eta=h=0$ into Eq.~\eqref{eq:H_XY}. 
Since the dynamics is unitary, the global state never relaxes, and the symmetry breaking in the initial state is retained throughout the time evolution. 
\par 
To see relaxation dynamics in such isolated systems, we consider a local subsystem $A$ of length $L_A$. 
The subsystem is described by the reduced density matrix $\rho_A=\Tr_{\bar{A}} [\ket{\psi(t)}\bra{\psi(t)}]$, where $\Tr_{\bar{A}}$ denotes the partial trace over the complement of $A$. 
The reduced state evolves non-unitarily and, in the thermodynamic limit $L,L_A\to\infty$ with $L_A/L\to0$, it relaxes to a generalised Gibbs ensemble of the XX Hamiltonian~\cite{rigol2007relaxation,barthel2008dephasing,cramer2008exact,calabrese2012quantum}.
Since the XX Hamiltonian preserves the U(1) spin-rotational symmetry around $z$ axis, the symmetry broken by the initial state is restored as the subsystem relaxes.  
The degree of symmetry breaking in $\rho_A(t)$ thus serves as a natural proxy for the relaxation in this system. 
\par 
The QFI is suited for describing the symmetry restoration within the subsystem, as it remains well-defined in the thermodynamic limit, where the relaxation occurs. 
The spin rotations around $z$ axis within subsystem $A$ are generated by the restricted magnetization $M_A=\sum_{i\in A}\sigma_i^z/2$. 
Using the Jordan-Wigner transformation~\cite{lieb1961two} and Wick's theorem~\cite{peschel2003calculation}, we obtain the exact result for the QFI with respect to $\rho_A$ and $M_A$ in the ballistic scaling limit $L_A,t\to \infty$ with $\zeta=t/L_A$ fixed (see Supplemental Information for the derivation): 
\begin{align}
    f_{\rho_A}(M_A)
    &=
    \lim_{\substack{L_A,t \to\infty \\ \zeta=t/L_A=\mathrm{const.}}}\qty(
    \frac{\lim_{L\to\infty}F_{\rho_A}(M_A)}{L_A})
    \nonumber 
    \\
    &=
    8 \int_{-\pi}^\pi 
    \frac{dk}{2\pi}
    x_k(\zeta)
    \abs{\bra{g}c_{-k}c_k\ket{g}}^2, 
    \label{eq:QFI_XX}
\end{align}
where $x_k(\zeta)=\max(1-2\zeta |v_k|,0)$ with $v_k=-\sin k$ being the group velocity of fermion.
\par 
Equation~\eqref{eq:QFI_XX} represents the celebrated quasiparticle picture for the symmetry restoration~\cite{rylands2024microscopic}: Since each Cooper pair consists of fermions with opposite momenta, they propagate in the opposite directions with velocities $\pm v_k$ after the quench. 
Consequently, the pairs gradually leave the subsystem, leading to the local symmetry restoration and, accordingly, the decrease in the QFI. 
Indeed, the weight function $x_k(\zeta)$ in Eq.~\eqref{eq:QFI_XX} precisely counts the number of such pairs that remain within the subsystem at the rescaled time $\zeta=t/L_A$. 
Therefore, the QFI corresponds to the number of Cooper pairs within the system of interest, even in the nonequilibrium regime. 
\par
\begin{figure}
    \raggedright
    \includegraphics[width=0.9\linewidth]{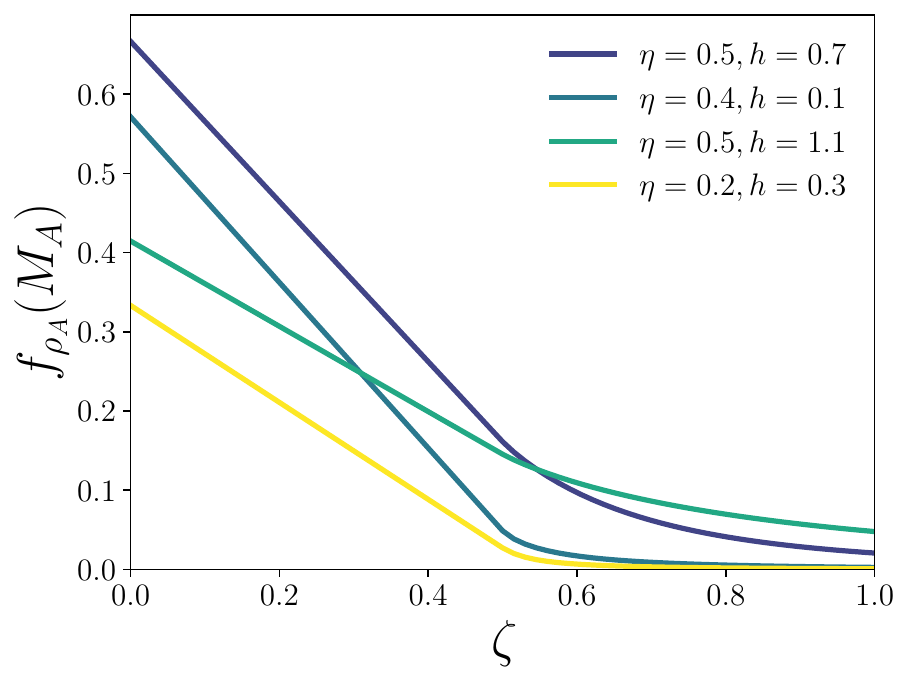}
    \caption{\textbf{Time evolution of the QFI for the XY spin chain quenched from finite transverse-field $h$ and anisotropy $\eta$ to $h=\eta=0$.}
    The curves plot the exact expression~\eqref{eq:QFI_XX} in the thermodynamic limit as a function of rescaled time $\zeta=t/L_A$. The intersections of the curves indicate the occurrence of the quantum Mpemba effect.}
    \label{fig:QFI_XY_quench}
\end{figure}

 
Figure~\ref{fig:QFI_XY_quench} shows Eq.~\eqref{eq:QFI_XX} for various initial states $\ket{g(h,\eta)}$ as a function of rescaled time $\zeta=t/L_A$. In several cases, curves of the QFI intersect at finite times, indicating that a state with larger initial symmetry breaking (larger QFI at $\zeta=0$) can exhibit faster local symmetry restoration than one closer to equilibrium. 
These intersections are the manifestation of the quantum Mpemba effect in the present setting. 
\par 
The quasiparticle picture naturally explains this effect. 
Given that the QFI corresponds to the number of Cooper pairs within the subsystem, fast-moving pairs accelerate the decay of the QFI as well as the symmetry restoration because they escape the subsystem quickly. 
In contrast, slow-moving pairs with $v_k\simeq 0$ ($k\simeq 0,\pi$) stay in the subsystem for longer periods, tending to preserve the initial symmetry breaking. 
Consequently, a state with more total Cooper pairs but fewer slow ones can relax faster than a state with fewer total pairs but more slow pairs, leading to the quantum Mpemba effect. 
While this mechanism has been previously elucidated through asymptotic analysis of entanglement asymmetry~\cite{rylands2024microscopic,murciano2024entanglement}, we provide the approximation-free derivation built upon the exact expression of the QFI in Eq.~\eqref{eq:QFI_XX}. 
\\~\\
\par\  \par \noindent{\bf \large Discussion}
\\
We have proposed the QFI as a faithful and generic measure of symmetry breaking, based on the resource theory of asymmetry.
By applying this quantity to a range of paradigmatic systems including BCS superconductors, topological spin chains, and non-equilibrium quantum quenches, we have demonstrated that the QFI faithfully captures symmetry-breaking phenomena across both equilibrium and non-equilibrium regimes. 
\par 
In addition to reproducing known results in rigourous way, our analysis uncovered several novel insights.
In superconductors, the QFI incorporates thermal decoherence through the purity of the state, thereby refining the description of symmetry breaking at finite temperature. 
In the XY spin chain, it displays a non-analytic behaviour at the topological phase transition where no local order parameter exists. 
In quench dynamics, it allows the approximation-free derivation of the conditions under which the quantum Mpemba effect occurs. 
\par 
From an experimental perspective, several state-of-the-art platforms offer promising opportunities to test the predictions presented here. Ultracold atomic gases~\cite{bloch2012quantum,gross2017quantum} and trapped-ion simulators~\cite{blatt2012quantum,lanyon2011universal}, in particular, allow precise control of interactions and geometry, making them ideal testbeds for probing QFI and its connection to symmetry breaking. 
Recent advances in quantum gas microscopy~\cite{schafer2020tools} and qubit-resolved measurements~\cite{brydges2019probing} in these systems enable the direct reconstruction of density matrices.
These techniques open the possibility of directly measuring QFI in both equilibrium and dynamical settings. 
Demonstrating the proposed connection between QFI and symmetry breaking in such experiments would not only provide a stringent test of our theoretical predictions but also open new avenues for dissecting quantum many-body phenomena from an information-theoric perspective. 
\\~\\
\noindent{\bf \large Methods}
\\
\noindent{\bf QFI of Gaussian states}
\\
We analytically calculate the QFI for BCS states, ground states of the XY spin chain, and time-evolved states for the quench dynamics starting from the XY to XX spin chain. These calculations exploit the fact that the states are Gaussian and thus obey Wick’s theorem. Here, we present a general formalism for computing QFI based on the Gaussianity of the states. Its specific applications to each example are provided in the Supplementary Information.
\par 
Let us denote by $f_i~(f_i^\dagger)~(i=1,2,...,n)$ the annihilation (creation) operator of a fermion on the $i$-th mode. If the state $\rho$ of the system is Gaussian and satisfies Wick’s theorem, then $\rho$ is uniquely characterized by the covariance matrix $\Sigma$~\cite{peschel2003calculation}, whose elements are defined as 
\begin{align}
    \Sigma_{ii'}=\frac{1}{2}
    \Tr\qty[\rho 
    \mqty([a_{i},a_{i'}] & [a_{i},b_{i'}] \\
    [b_{i},a_{i'}] & [b_{i},b_{i'}]
    )],
    \label{eq:cm}
\end{align}
where $a_{i}=f_i^\dag+f_i,~b_{i}=\im(f_i^\dag-f_i)$.
\par 
According to Eq.~\eqref{eq:QFI_Fidelity}, the QFI of $\rho$ with respect to the U(1) charge $Q = \sum_i f_i^\dag f_i$ can be calculated from the Uhlmann fidelity between $\rho$ and $\rho_\theta = e^{i \theta Q} \rho e^{-i \theta Q}$. Since $Q$ is quadratic in $f_i$ and $f_i^\dag$, $\rho_\theta$ is also Gaussian and is therefore fully characterized by its covariance matrix $\Sigma_\theta$, whose elements $[\Sigma_\theta]_{ii'}$ are obtained by replacing $\rho$ in Eq.~\eqref{eq:cm} with $\rho_\theta$. The Uhlmann fidelity between the two Gaussian states can be expressed in terms of their covariance matrices as~\cite{banchi2014quantum}.
\begin{multline}
    \mathcal{F}(\rho,\rho_\theta)
    =
    \det\Bigg[\sqrt{\frac{I-\Sigma}{2}} 
    \\+\sqrt{\frac{I-\Sigma}{2}}\sqrt{\sqrt{\frac{I+\Sigma}{I-\Sigma} \frac{I+\Sigma_\theta}{I-\Sigma_\theta} \sqrt{\frac{I+\Sigma}{I-\Sigma}}}}\Bigg]. 
\end{multline}
Evaluating the right-hand side of the above equation using the explicit forms of $\Sigma$ and $\Sigma_\theta$ for each example and substituting it into Eq.~\eqref{eq:QFI_Fidelity}, we calculate the QFI.
\\
\\~\\ 
\noindent {\bf \large Data availability}
\\
The data that support the plots within this paper are provided in the Source Data file.
\\~\\
\noindent {\bf \large Code availability}
\\
The computer codes used to generate the results that are reported in this paper are available from the authors upon reasonable request.
All correspondence and requests for materials can be addressed to any of the authors.
\\~\\
\noindent {\bf \large Acknowledgements}  
\\
SY was supprted by Grant-in-Aid for Young Scientists(Start-up) No.~25K23355. 
HT was supported by JSPS Grants-in-Aid for Scientific Research 
No. JP25K00924, and MEXT KAKENHI Grant-in-Aid for Transformative
Research Areas B ``Quantum Energy Innovation” Grant Numbers 24H00830 and 24H00831, JST MOONSHOT No. JPMJMS2061, and JST FOREST No. JPMJFR2365.
SE acknowledges support from JSPS KAKENHI Grant Numbers JP23H01174, JP25K00217, Matsuo Foundation, and Institute for Advanced Science, University of Electro-Communications.
\\~\\
\noindent {\bf \large Author Contributions}
\\
S. Y., S. E. and H. T. contributed to the numerical and analytic computations, the interpretation of the results, developing of the theory 
and to the writing of the manuscript.
\\~\\
\noindent {\bf \large Competing interests}
\\
The authors declare no competing interests.
\\~\\
\noindent{\bf \large References}
\bibliographystyle{apsrev4-2}
\bibliography{ref,RI_ref}

\clearpage
\newpage 
\setcounter{page}{1}
\onecolumngrid
\newcounter{equationSM}
\newcounter{figureSM}
\newcounter{tableSM}
\stepcounter{equationSM}
\setcounter{equation}{0}
\setcounter{figure}{0}
\setcounter{table}{0}
\setcounter{section}{0}
\makeatletter
\renewcommand{\theequation}{\textsc{sm}-\arabic{equation}}
\renewcommand{\thefigure}{\textsc{sm}-\arabic{figure}}
\renewcommand{\thetable}{\textsc{sm}-\arabic{table}}

\begin{center}
  {\large{\bf Supplemental Material}}
\end{center} 
\tableofcontents

\section{Minimal knowledge of Resource Theory and Resource Theory of Asymmetry}\label{SMsec:RT_and_RTA}
\subsection{Minimal knowledge of resource theory}\label{SMsubsec:RT}
\subsubsection{Framework}
Resource theory is a field that investigates the properties of “valuable” operations and states--those that do not belong to the set of free operations, $\calO_F$, which is defined as a subset of completely positive and trace preserving (CPTP) maps, and the set of free states, $\calS_F$, which is defined as a subset of quantum states \cite{Chitambar2019quantum}.
We consider the free operations as operations ``easy to implement” and the free states as states ``easy to prepare''.
Which operations and states are regarded as free depends on the physical setting to which the resource theory is applied, but the following is generally imposed as a minimal requirement:
\eq{
\calC(\rho)\in\calS_F,\enskip\forall\rho\in\calS_F,\enskip\forall\calC\in\calO_F.\label{resource1}
}
This condition states that “combining an easily prepared state with an easily implemented operation should not allow one to prepare a state that is not easy to prepare.”
There are many pairs $(\calS_F,\calO_F)$ satisfying this condition, and by replacing them, one obtains specific theories such as entanglement theory, quantum thermodynamics, and, as in this paper, the resource theory of asymmetry.

Once the free states and free operations are specified, one can define the “resource” from which the name “resource theory” derives. In a resource theory, a state that does not belong to $\calS_F$ is defined as a resource and is called a resource state. The reason for regarding such a state as a resource is that, when combined with free operations, it can enable phenomena that free operations alone cannot achieve. For example, in entanglement theory, using a Bell pair (a resource state) enables quantum teleportation, which can never be realized with LOCC (local operations and classical communication: free operation in the entanglement theory) and separable states (free states in the entanglement theory).

One of the main purposes of resource theory is to give proper ways of “measuring the amount of a resource.” We call a function of states $M$  a resource measure when $M$ satisfies the following properties:
\eq{
M(\rho)&\ge M(\calC(\rho)),\enskip\forall\calC\in\calO_F\label{resource2}\\
M(\rho)&\ge0\label{resource3}\\
\rho\in\calS_F&\Rightarrow M(\rho)=0.\label{resource4}
}
Here, when $\Leftarrow$ in \eqref{resource4} is satisfied, $M$ is called faithful.
Furthermore, when $M$ satisfies the following property, $M$ is called additive:
\eq{
M(\rho\otimes\sigma)=M(\rho)+M(\sigma).
}
This property directly implies
\eq{
\lim_{n\rightarrow\infty}\frac{M(\rho^{\otimes n})}{n}=M(\rho).
}
Therefore, the additivity of $M$ for product states naturally suggests its extensivity in the thermodynamic limit, where macroscopic systems can be approximated as composites of weakly correlated subsystems.

\subsubsection{``complete'' measure for asymptotic state conversion}

A variety of criteria can be considered for what constitutes a “good” resource measure, but the standard answer in resource theory is that if there exists a measure that determines the optimal asymptotic conversion rate $R(\rho\to\sigma)$ between the iid (independent and identically distributed) states defined as follows, then that measure is considered a good one:
\begin{definition}
Fix a set of free operations $\calO_F$ and a set of free states $\calS_F$.
Let $\rho$ and $\sigma$ be resource states, i.e., $\rho,\sigma\notin \calS_F$.
We say that $\rho$ is asymptotically convertible to $\sigma$ at rate $r$ for some real $r>0$ if
\begin{equation}
\exists \{ \mathcal{E}_N \} \mathrm{~s.t.~} \mathcal{E}_N\in\mathcal{O}_F,\enskip \lim_{N\to\infty}T
 \left( \mathcal{E}_N(\rho^{\otimes N}),\sigma^{\otimes \lfloor rN\rfloor} \right)=0,
\end{equation}
where $T(\cdot,\cdot)$ denotes the trace distance: $T(\rho,\sigma):=\frac{1}{2}\|\rho-\sigma\|_1$.
The optimal conversion rate $R(\rho\to\sigma)$ is defined as the supremum of $r$ for which the asymptotic conversion from $\rho$ to $\sigma$ is possible.
\end{definition}

If there exists a resource measure that completely determines this optimal conversion rate, it is called a complete resource measure \cite{sagawa_entropy_2022}. When referring simply to a “complete measure,” one may sometimes mean a set of measures giving necessary and sufficient conditions for single-shot state convertibility rather than for state copies as in the iid setting \cite{Liu2019}. Here, following Ref.~\cite{sagawa_entropy_2022}, we simply use the term complete measure, but for greater accuracy it might be better to call it a complete measure for asymptotic conversion.

For example, the entanglement entropy $E(\ket{\psi_{AB}}):=S(\Tr_B[\ketbra{\psi_{AB}}])$ (where $S(\rho):=-\Tr[\rho\log\rho]$ is the von Neumann entropy) completely determines the optimal conversion rate between pure states in the resource theory where $\calO_F$ consists of LOCC operations and $\calS_F$ is the set of separable states (i.e., entanglement theory) \cite{RToE1}:
\eq{
R(\ket{\psi_{AB}}\to\ket{\phi_{AB}})=\frac{E(\ket{\psi_{AB}})}{E(\ket{\phi_{AB}})}.
}

There are two main reasons why a complete measure is important.
First, by computing the measure alone, one can quantitatively evaluate “how much of one resource state can be converted into another / obtained from another.” This provides an intuitive understanding that the amount of the resource is characterized by the magnitude of the resource inherent in the state.
Second, the measure alone suffices to provide the ordering structure determined by the possibility or impossibility of iid asymptotic conversions. In particular, when the iid asymptotic conversion is given by a simple ratio, a total order structure analogous to thermodynamics emerges.

For these reasons, resource theory is guided by an empirical guideline that a complete measure, once identified, often turns out to be a physically fundamental quantity.
For example, the entanglement entropy and the nonequilibrium free energy serve as complete measures in entanglement theory and quantum thermodynamics, respectively.
The claim of this paper is that, the quantum Fisher information—which is a complete measure in the resource theory of asymmetry with respect to the $U(1)$ group—plays an essential role as an indicator of symmetry breaking in quantum many-body systems.

\subsection{Resource theory of asymmetry and QFI as resource measure}
\subsubsection{Framework of resource theory of asymmetry for $U(1)$-symmetry}
Here, we introduce the framework of resource theory of asymmetry.
In this paper, we focus on the case of $U(1)$-symmetry and $\mathbb{R}$-symmetry.

In a similar way as the other resource-theoretic frameworks, the resource theory of asymmetry specifies free states and free operations, which are called symmetric states and covariant operations, respectively.

Symmetric states are defined as follows.
Let $\rho$ denote a quantum state on system $S$ and $X_S$ the Hermitian operator representing the conserved quantity on $S$.
We say that $\rho$ is symmetric with respect to the unitary representation $\{e^{iX_St}\}$ if
\eq{
e^{iX_St}\rho e^{-iX_St} = \rho, \quad \forall t.
}
Equivalently, this condition holds precisely when $\rho$ commutes with $X_S$, i.e., $[\rho, X_S]=0$.
From this perspective, a symmetric state is simply one that exhibits no coherence (=non-diagonal elements) between eigenstates of the conserved quantity.

Covariant operations are defined as follows.
Consider a completely positive trace-preserving (CPTP) map $\calE_{S \to S'}$ between systems $S$ and $S'$, together with Hermitian operators $X_S$ and $X_{S'}$ representing the respective conserved quantities.
The operation $\calE_{S \to S'}$ is called covariant with respect to $\{e^{iX_St}\}$ and $\{e^{iX_{S'}t}\}$ when, for every $t$,
\begin{align}
\calE_{S \to S'}\left(e^{iX_St}(\cdot)e^{-iX_St}\right)
= e^{iX_{S'}t} \calE_{S \to S'}(\cdot)  e^{-iX_{S'}t}.
\label{covariantcond}
\end{align}

A notable fact about covariant operations is that they can always be implemented using a global unitary satisfying the conservation law together with an ancilla prepared in a free state.
To be concrete, given a covariant operation $\calE_{S \to S'}$ with respect to $\{e^{iX_St}\}$ and $\{e^{iX_{S'}t}\}$, one can always find ancillary systems $E$ and $E'$ (with $SE=S'E'$), conserved-quantity operators $X_E$ and $X_{E'}$ on $E$ and $E'$, a unitary $U$ on $SE$ and an ancilla state $\mu_E$ on $E$ such that \cite{Marvian_distillation}
\begin{align}
\calE_{S \to S'}(\cdot) &= \Tr_{E'}[U\left(\cdot \otimes \mu_E\right)U^\dagger],\\
U(X_S + X_E)U^\dagger &= X_{S'} + X_{E'},\\
[\mu_E,X_E]&=0
\end{align}

\subsubsection{Quantum Fisher information as a complete resource measure}

In the resource theory of asymmetry for the case of $U(1)$, the symmetric logarithmic derivative (SLD) quantum Fisher information for the state family generated by the unitary representation $\{e^{iXt}\}$ works as a complete measure.
First, we introduce the SLD Quantum Fisher information
\begin{definition}[SLD-Quantum Fisher information]
Let $\{\rho_t\}_{t\in \mathbb{R}}$ be a smooth state family.
For the family, the symmetric logarithmic derivative quantum Fisher information (SLD-QFI) is defined as follows:
\eq{
J_{\{\rho_t\}}|_{t=t_0}:=\langle L,L\rangle_{\rho_{t_0}},
}
where $\langle A,B\rangle_{\rho}:=\langle AB+BA\rangle _\rho/2$ and $L$ is the Hermitian operator satisfying
\eq{
\frac{\rho_{t_0}L+L\rho_{t_0}}{2}=\left.\frac{\partial \rho_t}{\partial t}\right|_{t=t_0}
}
\end{definition}
Using the SLD-QFI of the state family $\{\rho_t=e^{-i Xt}\rho e^{iXt}\}$, we can define a resource measure as:
\eq{
F_\rho(X):=J_{\{e^{-i Xt}\rho e^{iXt}\}}|_{t=0}.
}
We refer to $F_\rho(X)$ as QFI of $\rho$.
This $F_\rho(X)$ is a faithful measure and satisfies additivity for the tensor products:
\eq{
F_\rho(X)>0&\Leftrightarrow[\rho,X]\ne0,\\
F_{\rho\otimes\sigma}(X_{S_1}+X_{S_2})&=F_{\rho}(X_{S_1})+F_{\sigma}(X_{S_2}).
}
And furthermore, $F_\rho(X)$ is computable in the following two ways:
\eq{
 F_\rho(X)&=2 \sum_{i,j} \frac{(p_i-p_j)^2}{p_i+p_j} |\bra{i}X
\ket{j}|^2,\\
  F_\rho (X) &= -4 \partial_\theta^2 \mathcal{F}(\rho, e^{-i \theta X}\rho e^{i \theta X} )|_{\theta=0}, 
}
where $\{p_i,\ket{i}\}$ are the eigenvalues and eigenstates of $\rho$, and $\calF(\rho,\sigma):=\Tr[\sqrt{\sqrt{\rho}\sigma\sqrt{\rho}}]$ is the Uhlmann fidelity.

Furthermore, $F_\rho(X)$ is the complete measure for the asymptotic conversion between pure states:
\eq{
R(\ket{\psi}\rightarrow\ket{\phi})&=\frac{F_{\ket{\psi}}(X)}{F_{\ket{\phi}}(X)},
}
whenever $\tau(\ket{\psi},X)$ is an integer multiple of $\tau(\ket{\phi},X)$ (if not, $R(\ket{\psi}\rightarrow\ket{\phi})$=0), where $\tau(\rho,X)$ is 
\eq{
\tau(\rho,X):=\inf\{t|e^{-iXt}\rho e^{iXt}=\rho\}.
}
Furthermore, $F_\rho(X)$ is the complete measure for the asymptotic conversion from any pure state $\ket{\psi}$ to any mixed state $\rho$ whenever $\tau(\ket{\psi},X)$ is an integer multiple of $\tau(\rho,X)$ (if not, $R(\ket{\psi}\rightarrow\rho)$=0):
\eq{
R(\ket{\psi}\rightarrow\rho)&=\frac{F_{\ket{\psi}}(X)}{F_{\rho}(X)},
}

\section{QFI of the BCS state}\label{SMsec:BCS}
Here we give a detailed derivation of Eq.~\eqref{eq:QFI_BCS_finite_T} in the main text. 
We first note that the Gibbs state $\rho \propto e^{-H_\mathrm{BCS}/T}$ factorizes in the momentum basis as $\rho_T =\bigotimes_\mathbf{k}\rho_{T,\mathbf{k}}$, where
\begin{gather}
    \rho_{T,\mathbf{k}}=\frac{e^{-H_\mathbf{k}/T}}{\Tr e^{-H_\mathbf{k}/T}}, 
    \\
    H_\mathbf{k} = (\xi_\mathbf{k}-\mu)N_\mathbf{k} 
    -\Delta 
    (c_{\mathbf{k}\uparrow}^\dag c_{\mathbf{-k}\downarrow}^\dag +\mathrm{H.c.}),
    \label{eq:H_k}
\end{gather} 
with $N_\mathbf{k}=c_{\mathbf{k}\uparrow}^\dag c_{\mathbf{k}\uparrow}+c_{\mathbf{-k}\downarrow}^\dag c_{\mathbf{-k}\downarrow}$. 
This factorization and the additivity of the QFI allow us to decompose $F_{\rho_T}(N)$ as 
\begin{align}
    F_{\rho_T}(N) = \sum_\mathbf{k} F_{\rho_{T,\mathbf{k}}}(N_\mathbf{k}). 
\end{align}
Using the expression~\eqref{eq:QFI_Fidelity} in the main text, $F_{\rho_{T,\mathbf{k}}}(N_\mathbf{k})$ in the above equation can be written as 
\begin{align}
    F_{\rho_{T,\mathbf{k}}}(N_\mathbf{k})
    = 
    -4 \partial_\theta^2 \mathcal{F}
    \qty(\rho_{T,\mathbf{k}},e^{-\im \theta N_\mathbf{k}}\rho_{T,\mathbf{k}} e^{\im \theta N_\mathbf{k}})\,|_{\theta=0}. 
    \label{eq:QFI_BCS_fidelity}
\end{align}
\par 
Now, the problem reduces to how to calculate the Uhlmann fidelity between $\rho_{T,\mathbf{k}}$ and $e^{-\im \theta N_\mathbf{k}}\rho_{T,\mathbf{k}} e^{\im \theta N_\mathbf{k}}$. 
Since $H_\mathbf{k}$ and $N_\mathbf{k}$ are quadratic operators, its thermal state $\rho_{T,\mathbf{k}}$ and $e^{-\im \theta N_\mathbf{k}}\rho_{T,\mathbf{k}}e^{\im \theta N_\mathbf{k}}$ satisfy Wick's theorem and are fully described by the covariance matrix defined by~\cite{peschel2003calculation} 
\begin{align}
    \Sigma_{T,\mathbf{k}}(\theta)
    =
    \mqty(
    \Sigma_{T,\mathbf{k}}^{\uparrow \uparrow}(\theta)
    &
    \Sigma_{T,\mathbf{k}}^{\uparrow \downarrow}(\theta)
    \\
    \Sigma_{T,\mathbf{k}}^{\downarrow \uparrow}(\theta)
    &
    \Sigma_{T,\mathbf{k}}^{\downarrow \downarrow}(\theta)
    ),
    \label{eq:covariance matrix}
\end{align}
where 
\begin{align}
    \Sigma_{T,\mathbf{k}}^{\sigma\sigma'}(\theta)
    =
    \frac{1}{2}
    \Tr\qty[ e^{-\im \theta N_\mathbf{k}}\rho_{T,\mathbf{k}}e^{\im \theta N_\mathbf{k}}
    \mqty(
    [a_{\mathbf{k}\sigma},a_{\mathbf{k}\sigma'}]
    & 
    [a_{\mathbf{k}\sigma},b_{\mathbf{k}\sigma'}]
    \\
    [b_{\mathbf{k}\sigma},a_{\mathbf{k}\sigma'}]
    & 
    [b_{\mathbf{k}\sigma},b_{\mathbf{k}\sigma'}]
    )
    ].
\end{align}
Here, we introduced the Majorana fermion operators defined as 
\begin{align}
    a_{\mathbf{k}\sigma}
    &=
    \begin{cases}
        c_{\mathbf{k}\uparrow}+c_{\mathbf{k}\uparrow}^\dag & \sigma=\uparrow\\
        c_{-\mathbf{k}\downarrow}+c_{-\mathbf{k}\downarrow}^\dag & \sigma=\downarrow\\
    \end{cases},
    \\
    b_{\mathbf{k}\sigma}
    &=
    \begin{cases}
        \im(c_{\mathbf{k}\uparrow}-c_{\mathbf{k}\uparrow}^\dag) & \sigma=\uparrow\\
        \im(c_{-\mathbf{k}\downarrow}-c_{-\mathbf{k}\downarrow}^\dag) & \sigma=\downarrow\\
    \end{cases}.
\end{align}
Evaluating each element in Eq.~\eqref{eq:covariance matrix}, one finds 
\begin{align}
    \Sigma_{T,\mathbf{k}}(\theta)
    =
    [1-\Tr(\rho_{T,\mathbf{k}}N_\mathbf{k})]
    I\otimes \sigma^y 
    +
    2\Re[\Tr(\rho_{T,\mathbf{k}}c_{-\mathbf{k}}c_{\mathbf{k}\uparrow})]
    \sigma^y \otimes \sigma^xe^{2\im \theta \sigma^y}
    +
    2\Im[\Tr(\rho_{T,\mathbf{k}}c_{-\mathbf{k}})]
    \sigma^y \otimes \sigma^z e^{2\im \theta \sigma^y}. 
    \label{eq:BCS_CM}
\end{align}
The Uhlmann fidelity between fermionic Gaussian states can be expressed in terms of their covariance matrix as~\cite{banchi2014quantum} 
\begin{align}
    \mathcal{F}\qty(\rho_{T,\mathbf{k}},e^{-\im \theta N_\mathbf{k}} \rho_{T,\mathbf{k}} e^{\im \theta N_\mathbf{k}})
    =
    \det[\frac{I-\Sigma_{T,\mathbf{k}}(0)}{2}]^{\frac{1}{2}}
    \det[I+\sqrt{
    \sqrt{\frac{I+\Sigma_{T,\mathbf{k}}(0)}{I-\Sigma_{T,\mathbf{k}}(0)}}
    \frac{I+\Sigma_{T,\mathbf{k}}(\theta)}{I-\Sigma_{T,\mathbf{k}}(\theta)}
    \sqrt{\frac{I+\Sigma_{T,\mathbf{k}}(0)}{I-\Sigma_{T,\mathbf{k}}(0)}}
    }].
    \label{eq:BCS_fidelity}
\end{align}
Substituting Eq.~\eqref{eq:BCS_CM} into Eq.~\eqref{eq:BCS_fidelity} and expanding it in terms of $\theta$, we obtain 
\begin{align}
    \mathcal{F}\qty(\rho_{T,\mathbf{k}},e^{-\im \theta N_\mathbf{k}} \rho_{T,\mathbf{k}}e^{\im \theta N_\mathbf{k}} )
    = 
    1-
    \frac{2|\Tr(\rho_{T,\mathbf{k}}c_{-\mathbf{k}\downarrow}c_{\mathbf{k}\uparrow})|^2}{\sqrt{P_{T,\mathbf{k}}}}
    \theta^2
    +O(\theta^3), 
\end{align}
where $P_{T,\mathbf{k}}=\Tr(\rho_{T,\mathbf{k}}^2)$. 
Inserting it into Eq.~\eqref{eq:QFI_BCS_fidelity} and taking the summation over $\mathbf{k}$, we arrive at Eq.~\eqref{eq:QFI_BCS_finite_T}. 
\par 
To calculate Eq.~\eqref{eq:QFI_BCS_finite_T} in the main text numerically, we replace the summation over $\mathbf{k}$ with an integral over $\epsilon=\xi_\mathbf{k}-\mu$ by introducing the density of states $D(\epsilon)$ and energy cutoff $\omega_\mathrm{D}$. 
It results in 
\begin{align}
    F_{\rho_T}(N)
    =
    16\Delta^2 D_\mathrm{F}
    \int_0^{\omega_\mathrm{D}} d\epsilon
    \frac{ \tanh^2\qty(\frac{\epsilon^2+\Delta^2}{2T})}{(\epsilon^2+\Delta^2)\qty[
    1+\tanh^2\qty(\frac{\epsilon^2+\Delta^2}{2T})]}. 
    \label{eq:QFI_BCS_integral}
\end{align}
Here, we approximated $D(\epsilon)=D$ as its value at the Fermi level $D_\mathrm{F}=D(0)$. 
In the same way, the gap equation $\Delta = U \sum_\mathbf{k} \langle c_{-\mathbf{k}\downarrow} c_{\mathbf{k}\uparrow}\rangle$ reduces to 
\begin{align}
    1=UD_\mathrm{F} \int_0^{\omega_\mathrm{D}} d\epsilon
    \frac{ \tanh(\frac{\sqrt{\epsilon^2+\Delta^2}}{2T})}{\sqrt{\epsilon^2+\Delta^2}}.
    \label{eq:gap equation}
\end{align}
Numerically calculating Eqs.~\eqref{eq:QFI_BCS_integral} and \eqref{eq:gap equation}, we obtain the curves in Fig.~\ref{fig:QFI_BCS}.

\section{QFI in the quench dynamics of XY spin chain}\label{SMsec:quench}

Here, we give the derivation of Eq.~\eqref{eq:QFI_XX}. Performing the Jordan-Wigner transformation, the XY spin chain is mapped to the spinless free fermion system and, therefore, the reduced density matrix can be fully characterised by the covariance matrix, as in the case of the BCS theory. 
In the present case, the covariance matrix of the reduced density matrix is defined in terms of spinless free fermions as 
\begin{align}
    \Sigma_{ii'}
    = 
    \frac{1}{2}
    \left\langle 
    \mqty(
    [a_i,a_{i'}] & [a_i,b_{i'}] \\ [b_i,a_{i'}] & [b_i,b_{i'}]
    )
    \right\rangle,\quad i,i'\in [1,L_A],
\end{align}
where $a_i=c_i+c_i^\dag$ and $b_i=\im(c_i^\dag-c_i)$. Evaluating each element of $\Sigma$ with the time evolved state $\ket{\psi(t)}=e^{-\im t H_\mathrm{XX}}\ket{g}$ and taking the thermodynamic limit $L\to\infty$, we obtain 
\begin{align}
    \Sigma_{ii'}
    = 
    \int_{-\pi}^{\pi}
    \frac{dk}{2\pi}
    e^{-\im k(i-i')}s_k(t), 
    \label{eq:Sigma_XX}
\end{align}
where 
\begin{align}
    s_k(t)
    = 
    (2n_k-1)\sigma^y
    +
    2(\Im[m_k] \sigma^x 
    +\Re[m_k] \sigma^z )e^{2\im t \cos k \sigma_y}, 
\end{align}
with $n_k = \bra{g}c_k^\dag c_k\ket{g}$ and $m_k =\bra{g}c_{-k}c_k \ket{g}$. 
In the same way, we obtain the covariance matrix $\Sigma_\theta$ of $\rho_{A,\theta}=e^{-\im \theta M_A}\rho_Ae^{\im \theta M_A}$ as 
\begin{align}
    \Sigma_\theta
    =
    e^{\im \theta I\otimes \sigma^y} \Sigma e^{-\im \theta I \otimes \sigma^y}. 
\end{align}
\par 
The Uhlmann fidelity between $\rho_A$ and $\rho_{A,\theta}$ can be written in terms of their covariance matrices by simply replacing $\Sigma_{T,\mathbf{k}}(0)$ and $\Sigma_{T,\mathbf{k}}(\theta)$ in Eq.~\eqref{eq:BCS_fidelity} with $\Sigma$ and $\Sigma_\theta$. 
That is, 
\begin{align}
    \ln(\mathcal{F}(\rho_A,\rho_{A,\theta}))
    =
    \Tr\log \Bigg(
    \qty[\frac{I-\Sigma}{2}]^{1/2}
    \Bigg[ 
    I+\sqrt{
    \sqrt{\frac{I+\Sigma}{I-\Sigma}}
    \frac{I+\Sigma_\theta}{I-\Sigma_\theta}
    \sqrt{\frac{I+\Sigma}{I-\Sigma}}
    }\Bigg]\Bigg)
    .
\end{align}
To derive the QFI in the ballistic scaling limit $L_A,t\to\infty$ with $\zeta=t/L_A$ fixed, we expand the right-hand side of the above equation in terms of the moments of $\Sigma$ as 
\begin{align}
    \ln(\mathcal{F}(\rho_A,\rho_{A,\theta}))=\sum_{n,\mathbf{a},\mathbf{b}}
    \mathcal{M}_{\mathbf{a},\mathbf{b}}(\theta)
    \Tr~[\Sigma^{a_1}(I\otimes\sigma^y)^{b_1}\Sigma^{a_2}...
    \Sigma^{a_{n-1}}(I\otimes\sigma^y)^{b_{n-1}}\Sigma^{a_n}(I\otimes \sigma^y)^{b_n}]. 
    \label{eq:M_expand}
\end{align}
Here, $\mathbf{a}$ and $\mathbf{b}$ are the $n$-dimensional vectors whose elements are integers and the expansion coefficients $\mathcal{M}_\mathbf{a,b}(\theta)$ is determined so that 
\begin{align}
\sum_{n,\mathbf{a,b}}\mathcal{M}_{\mathbf{a},\mathbf{b}}(\theta)
    A^{a_1}B^{b_1}...
    =
    \log \Bigg(
    \qty[\frac{I-A}{2}]^{1/2}
    \Bigg[I+
    \Bigg(\sqrt{\frac{I+A}{I-A}}
    \frac{I+e^{-\im \theta B}Ae^{\im \theta B}}{I-e^{-\im \theta B}Ae^{\im \theta B}}
    \sqrt{\sqrt{\frac{I+A}{I-A}}}\Bigg)^{1/2}
    \Bigg]
    \Bigg), 
    \label{eq:M_ab}
\end{align}
for matrices $A$ and $B$. 
Substituting Eq.~\eqref{eq:Sigma_XX} into Eq.~\eqref{eq:M_expand}, we obtain 
\begin{align}
    \ln(\mathcal{F}(\rho_A,\rho_{A,\theta}))
    =
    \sum_{n,\mathbf{a,b}}\mathcal{M}_{\mathbf{a,b}}(\theta)
    \int \limits_{[-\pi,\pi]^{M}}
    \frac{d^M\boldsymbol{k}}{(2\pi)^M}
    \sum_{i_1=1}^{L_A}\cdots \sum_{i_M=1}^{L_A}
    e^{-\im \sum_{j=1}^m k_j(i_{j}-i_{j+1})}
    m_{\mathbf{a,b}}(\boldsymbol{k}), 
    \label{eq:lnF_1}
\end{align}
where $M=\sum_{i=1}^n a_i$ and 
\begin{align}
    m_{\mathbf{a,b}}(\boldsymbol{k})
    =
    \Tr[s_{k_1}s_{k_2}...s_{k_{a_1}} (\sigma^y)^{b_1}s_{k_{a_1}+1}...s_{k_{a_1+a_2}}(\sigma^y)^{b_2}...
    s_{k_{a_1+...+a_n}}(\sigma^y)^{b_n}].
\end{align}
Using the identity 
\begin{align}
    \sum_{i=1}^\ell e^{-\im k\ell}
    =
    \frac{L_A}{2}
    \int_{-1}^1 
    du \frac{k}{\sin(k/2)}e^{-\im \frac{L_A}{2}ku},
\end{align}
we replace the summations over $i_{1\sim M}$ in Eq.~\eqref{eq:lnF_1} with the integrals over $u_{1\sim M}$ as 
\begin{align}
    \ln(\mathcal{F}(\rho_A,\rho_{A,\theta}))
    =
    \sum_{n,\mathbf{a,b}}
    \mathcal{M}_{\bf a,b}(\theta)
    \frac{L_A^M}{(4\pi)^M}
    \int \limits_{[-\pi,\pi]^M} \!\!\!\!\!\!
    d^M \boldsymbol{k}
    \int \limits_{[-1,1]^M}\!\!\!\!\!\!
    d^M \boldsymbol{u}
    e^{\frac{\im}{2}L_A\sum_{i=1}^M k_i(u_{i+1}-u_i)}
    m_{\mathbf{a,b}}(\boldsymbol{k}) C(\boldsymbol{k}), 
    \label{eq:lnF_2}
\end{align}
with $C(\boldsymbol{k})=\prod_{i=1}^M \frac{k_{i}-k_{i+1}}{2\sin([k_i-k_{i+1}]/2)}$. 
Decomposing $s_k$ as 
\begin{align}
    s_k = \bar{s}_k +s_k^+ e^{-2\im t \cos k} + s_k^- e^{2\im t \cos k},
\end{align}
with 
\begin{gather}
    \bar{s}_k = (2n_k-1)\sigma^y, 
    \\
    s_k^\pm = (\Im[m_k]\pm\im \Re[m_k])(\sigma^x\mp \im \sigma^z),
\end{gather}
$m_\mathbf{a,b}(\boldsymbol{k})$ in Eq.~\eqref{eq:lnF_2} can be expanded as~\cite{caceffo2024entangled}
\begin{align}
    m_{\mathbf{a,b}}(\boldsymbol{k})
    =
    \sum_{p=0}^{\lfloor \frac{M}{2} \rfloor}
    \sum_{1\leq j_1<j_2...<j_{2p}\leq M} 
    \Tilde{m}_{\mathbf{a,b}}(\boldsymbol{k},\qty{j_1,...,j_{2p}})
    \exp(-2\im t \sum_{i=1}^{2p}(-1)^i \cos(k_{j_i})). 
    \label{eq:m_decomposition}
\end{align}
Here, $\Tilde{m}_{\bf a,b}(\boldsymbol{k},\qty{j_1,...,j_{2p}})$ is obtained by replacing $s_{k_{2j-1}}$, $s_{k_{2j}}$, and $s_{k_{i\notin\qty{j_1,...,j_{2p}}}}$ in $m_{\mathbf{a,b}}(\boldsymbol{k})$ with $s_{k_{2j-1}}^+$, $s_{k_{2j}}^-$, and $1+\bar{s}_k+\sigma^y$, respectively. 
Substituting Eq.~\eqref{eq:m_decomposition} into Eq.~\eqref{eq:lnF_2}, we obtain 
\begin{multline}
    \ln(\mathcal{F}(\rho_A,\rho_{A,\theta}))
    =
    \sum_{n,\mathbf{a,b}}
    \mathcal{M}_{\bf a,b}(\theta)
    \frac{\ell^M}{(4\pi)^M}
    \int \limits_{[-\pi,\pi]^M} \!\!\!\!\!\!
    d^M \boldsymbol{k}
    \int \limits_{[-1,1]^M}\!\!\!\!\!\!
    d^M \boldsymbol{u}
    \sum_{p=0}^{\lfloor \frac{M}{2}\rfloor}
    \sum_{1\leq j_1<...<j_{2p}\leq M}
    C(\boldsymbol{k})\Tilde{m}_{\mathbf{a,b}}(\boldsymbol{k},\qty{j_1,...,j_{2p}})
    \\
    e^{\im \frac{L_A}{2} \qty[\sum_{i=1}^M k_i(u_{i+1}-u_i) -4 \zeta \sum_{i=1}^{2p}(-1)^i \cos(k_{j_i})]}. 
    \label{eq:lnF_3}
\end{multline}
\par 
Using the fact that the phase factor of the integrand of Eq.~\eqref{eq:lnF_3} is proportional to $L_A$, we can exactly calculate leading contribution in the limit $L_A\to \infty$ to the integral over $k_{2\sim M}$ and $u_{1\sim M}$ by applying the multi-dimensional stationary phase approximation~\cite{wong2001asymptotic}. 
The result is 
\begin{multline}
    \ln(\mathcal{F}(\rho_A,\rho_{A,\theta}))
    =
    L_A \sum_{n,\mathbf{a,b}}M_{\mathbf{a,b}}(\theta)
    \int_{-\pi}^\pi 
    \frac{dk}{8\pi}
    \Bigg\{
    x_k(\zeta)
    \Tr\qty[\prod_{i=1}^n (\bar{s}_k + s_k^+ + s_k^-)^{a_i} (\sigma^y)^{b_i}]
    \\
    +(1-x_k(\zeta)) \Tr\qty[\prod_{i=1}^n (\bar{s}_k)^{a_i}(\sigma^y)^{b_i}]
    \Bigg\}
    +O(L_A^{1/2})
    \label{eq:lnF_4}
\end{multline}
Taking the summations over $\mathbf{a}$ and $\mathbf{b}$ in Eq.~\eqref{eq:lnF_4} using Eq.~\eqref{eq:M_ab} and then evaluating the trace, we obtain 
\begin{align}
    \ln(\mathcal{F}(\rho_A,\rho_{A,\theta}))
   =
    L_A 
    \int_{-\pi}^\pi
    \frac{dk}{8\pi}x_k(\zeta)\ln(1-4|m_k|^2 \sin^2 \theta)
    +O(L_A^{1/2}). 
    \label{eq:lnF_5}
\end{align}
Exponentiating the both sides of Eq.~\eqref{eq:lnF_5} and taking the second derivative with respect to $\theta$, one finds 
\begin{align}
    \frac{F_{\rho_A}(M_A)}{L_A}
    =
    -\frac{4}{L_A}\partial_\theta^2 \mathcal{F}(\rho_A,\rho_{A,\theta})|_{\theta=0}
    = 
    8 \int_{-\pi}^\pi \frac{dk}{2\pi} x_k(\zeta) |m_k|^2
    +O(L_A^{-1/2}). \label{eq:QFI_XX_}
\end{align}
Taking the thermodynamic limit $L_A\to\infty$ in Eq.~\eqref{eq:QFI_XX_} and recalling that $|m_k|=\bra{g}c_{-k}c_k \ket{g}$, we finally arrive at Eq.~\eqref{eq:QFI_XX} in the main text.

\end{document}